\documentclass{aa}
\usepackage{epsfig}

\newcommand{\mic}{\,{\rm \mu m} } 

\begin{document}

\title{FIRBACK II. Data Reduction and Calibration of the 170 $\mu m$ ISO\thanks{Based on observations with ISO, an ESA project with instruments
funded by ESA Member States (especially the PI countries: France,
Germany, the Netherlands and the United Kingdom) and with the
participation of ISAS and NASA.} Deep Cosmological Survey} 

\titlerunning{FIRBACK II. Data Reduction and Calibration}

\author{Guilaine Lagache  
	\inst{1} \and
	Herv\'e Dole 
        \inst{2,1}}

\institute{
Institut d'Astrophysique Spatiale, B\^at.  121, Universit\'e Paris XI, 91405 Orsay Cedex, France \and
Steward Observatory, University of Arizona, 933 N Cherry Ave, Tucson, AZ, 85721, USA}

\offprints{Guilaine.Lagache@ias.u-psud.fr}

\date{Received 24 January 2001; Accepted 5 March 2001}  
          
\abstract{We present the final reduction and calibration
of the FIRBACK ISOPHOT data. FIRBACK is a deep cosmological
survey performed at 170 $\mu$m. This paper 
deals with the ISOPHOT C200 camera with the C160 filter.
We review the whole data reduction process and
compare our final calibration with DIRBE 
(for the extended emission) and IRAS (for point sources).
The FIRBACK source extraction and galaxy counts
is discussed in a companion paper
(\cite{Doleprep}).}

\maketitle

\section{Introduction}
With more than 150 hours of observations, FIRBACK (Far-InfraRed BACKground) 
is one of the largest observational programmes made with the
ISOPHOT instrument on board the Infrared Space Observatory (ISO) 
satellite. This cosmological deep survey at 170 $\mu$m
covers more than 4 square degrees located in
two Northern and one Southern fields (Lagache 1998, Dole 2000a).
There are 
106 sources detected above the sensitivity limit 
 (180 mJy, 4$\sigma$). 
The number of sources detected above 135 mJy ($ 3 \sigma $) is 196. 
The first result of this survey is the high number of sources observed when compared to 
no, or moderate, evolution models for infrared galaxies (e.g. Dole 2000a). 
Preliminary results in the so-called Marano1 field were published in
Puget et al. (1999), and on the whole survey by Dole et al. (2000b).
FIRBACK also allows for the first time the detection of the
CFIRB fluctuations (Lagache \& Puget 2000; Puget \& Lagache, 2001).\\

After a presentation of the observational
issues of the FIRBACK survey, we present in this paper
the final stage of the data reduction
and calibration. We use the Phot Interactive
Analysis (PIA, Gabriel et al. 1997) for the standard reduction
and calibration (Sect. \ref{PIA}) with some extra developments:
flat-fielding, transient corrections and reprojection (Sect. 3.2-3.7).
In Sect. \ref{cal}, we review our final map calibration
and compare it with absolute photometric calibration from other instruments. 
We then conclude in Sect. 5.

\begin{table}
\caption{FIRBACK fields main characteristics}
\label{FIRBACK_def}
\begin{tabular}{|l|c|c|c|c|c|} \hline 
Field & l & b & surface & t$_{int}$ & Number of\\ 
& deg & deg & deg$^2$ & sec & rasters \\ \hline
FSM & 270 & -52 & 0.95 & 256 & 2$\times$11 \\ \hline
FN1 & 84 & +45 & 1.98 & 128 & 2$\times$9 \\ \hline
FN2 & 65 & +42 & 0.96 & 128 & 4$\times$4 \\ \hline 
\end{tabular}\\
\end{table}

\section{FIRBACK observations}
FIRBACK covers about 4 square degrees in 3 high galactic latitude fields (chosen to have 
very low HI column density, typically $N_{HI} \le 10^{20} cm^{-2}$), the so-called
N1 (FN1), N2 (FN2) and Marano (FSM) fields.
The observations were performed at an effective wavelength of 
170$\mic$ with the C200 array (2x2 pixels) of ISOPHOT. The pixel field of view is 1.5 arcmin.
We use the AOT (Astronomical Observation Template) PHT22 in the multi-pointing staring raster mode. 
Each FIRBACK field (Table \ref{FIRBACK_def}) consists of several rasters of 17$\times$17
pixels{\footnote{Except for the FSM1 field (Puget et al. 1999) 
where the rasters are 19$\times$19 pixels}}: 
\begin{itemize}
\item Eleven rasters were done in the so-called FN1 field.
These rasters were re-observed with a shift in position 
that corresponds to a fraction of an ISOPHOT pixel to provide proper sampling where possible.
Observations were done from the ISO revolution 753 to 774.
This field covers the same area as the ELAIS N1 15 $\mu$m and 90 $\mu$m observations
(Oliver et al. 2000). 
\item Nine rasters were done in the so-called FN2 field. 
These rasters were again re-observed with a shift in position.
The field covers the same area as the ELAIS N2 15 $\mu$m and 90 $\mu$m observations.
These data are FIRBACK/ELAIS observations. They 
were done from the ISO revolution 785 to 798.
\item  One raster was observed in the so-called FSM1 field (formerly called
Marano 1 field) during the revolution 593 and 
three rasters in the so-called FSM234 field (from revolution 739 to 744). Rasters
have been observed four times. Displacements between the four independent observations 
correspond to about 0.5 pixel, except for the FSM1 field where the displacements 
correspond to 2 pixels
(note that the overlapped surface with the ``original'' Marano field
is less than 15$\%$).
\end{itemize}

Each raster is performed in the spacecraft (Y, Z) coordinate system which
is parallel to the edges of the detector array, with one pixel overlap in
both Y and Z direction. The exposure time is 16s per pixel
and thus 128, 128 and 256s per sky position in the FN1, FN2 and
in the FSM fields respectively.\\

We also have a PHT25 measurement in the FSM1 field. This AOT is the absolute
photometry mode for PHT-C, in which photometric calibration is achieved
by chopping against the internal fine calibration source. For low fluxes
(i.e. in our case), chopping is also done against the switched-off ``Fine Calibration Source'', which
has a temperature level of about 4K. This is the temperature of the optical
support on which the instrument is mounted. Such a measurement serves
to define a zero point. This mode is especially well suited
for observations aiming at accurate determination of the absolute
brightness of the background emission.

\section{Data reduction}

\subsection{\label{PIA}PIA data reduction and calibration} 
We use PIA, the ISOPHOT Interactive Analysis software 
version 7.2.2 (Gabriel et
al. 1997), to correct for instrumental effects, the glitches induced
by cosmic particles and to provide an initial calibration. First we apply
the non-linearity correction due to 2 independent effects: the 
non-linearity of the Cold Readout Electronics and the downward curving
ramp due to de-biasing. Deglitching is performed for each individual
ramp and then the mean signal per position is derived by averaging
the ramp slopes (we do not apply any transient correction
at this stage). We then apply a second deglitching (for every chopper
position, a ``running mean'' method is applied; signals that are far
away by a number of sigma, typically 5, are flagged).
The dark current, which represents less than 5$\%$ of the signal, 
is subtracted using the orbit dependent calibration files. 
We also apply the
``reset interval correction'', a correction which represents
less than 6$\%$ of the signal.\\

We do exactly the same data reduction for the two ``Fine Calibration Source''
(FCS) measurements. The FCS measurements before and after
the observation vary by less than 3$\%$ for most of the rasters.
We decide to perform the first calibration (from V/s to MJy/sr) 
by deriving the mean value of the two FCS measurements. We prefer not 
to use the interpolation between the two FCSs when they are different
since it may induce Long Term Transients (LTT) that are not
necessary real. On the contrary, when real, the
interpolation bewteen the two FCSs could correct for the LTT, but we
prefer to correct the LTT following Sect.3.3.\\

\begin{figure}
	\begin{center}
	\epsfxsize=9cm
	\epsfbox{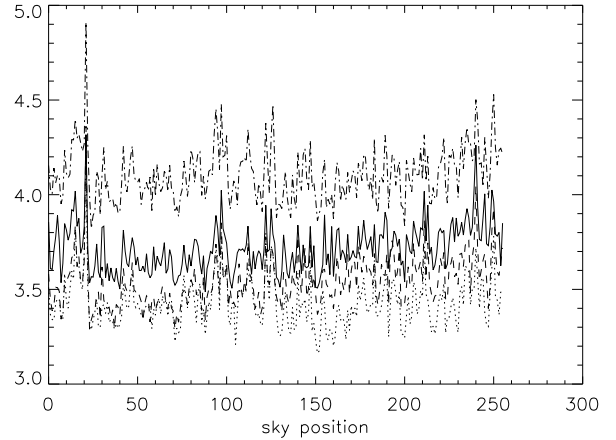}
	\caption{\label{signal_exam} Signal in MJy/sr of the four pixels as a function
of the sky position for one raster. The mean signal level is not the same (no flat-field 
correction has been applied at this stage) but the signal variations are
highly reproductible.}
	\end{center}
\end{figure}

At this stage of the data reduction, we have the signal variation
(in MJy/sr) as a function of time (or position) for each pixel 
in each raster. We see from Fig. \ref{signal_exam} 
that the instrumental noise is very low. We have, per sky position, 
a very high reproductibility of the data.

\subsection{The PHT25 measurement}
\begin{figure}
\epsfxsize=9.cm
\epsfbox{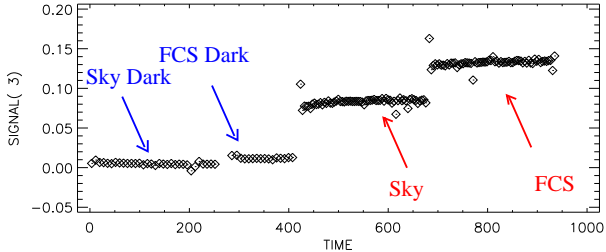}
\vspace{-3.5cm}
\caption{\label{pht25} Absolute PHT25 measurement sequence for one pixel
(in V/s)}
\end{figure}

The AOT PHT25 absolute measurement in the FSM1 field has the following
characteristics (Fig. \ref{pht25}):
\begin{itemize}
\item 256s for the ``sky dark'': observation of the cold opaque filter
\item 128s for the ``FCS dark'': observation of the cold FCS (turned off)
\item 256s on sky
\item 256s on the FCS
\end{itemize}

With this measurement we check that:
\begin{itemize}
\item The FCS value after 256s and the one obtained with 32s 
integration time (which is the integration time in the PHT22 observations)
differs by less than 4$\%$
\item The difference between the observed dark current (``sky dark'') and
the default PIA one is less than 10$\%$
\item The difference between the FCS dark current (``FCS dark'') and 
the default PIA one is less than 2$\%$
\end{itemize}

We obtain for the sky measurement a brightness{\footnote{We correct the brightness
for the flat-field using the average values given in Sect. 3.4 and
apply the extended calibration factor correction (Sect 4.3).}} of 3.10$\pm$0.14 MJy/sr,
to be compared to the value measured on the final map (Fig. 6), which is 3.08 MJy/sr.
We have a very good agreement between both measurements;
we do not observe any significant discrepancy
between the ``raster mode'' observations and the ``absolute mode''
one.

\subsection{\label{LTT}Long term transients}
\begin{figure}
\epsfxsize=9.cm
\epsfbox{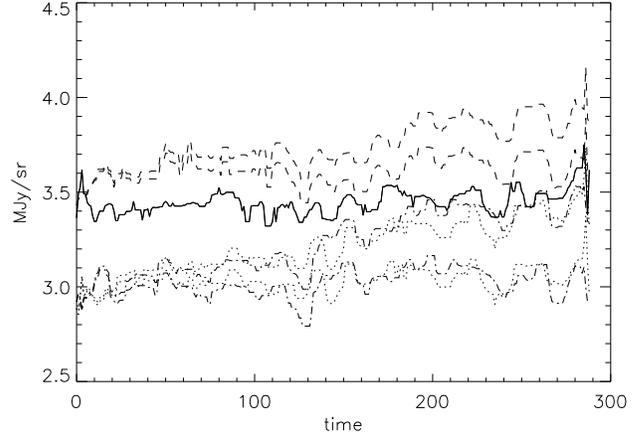}
\caption{\label{LTT_1} Illustration of the LTT correction on a 17$\times$17
raster. The thick solid line is the pixel signal (median
filtered over 9 position) with the smallest slope.
The long term variation of the three other pixel signals (thin lines:
dot, dot-dash, dash) are corrected 
in order to follow the same slope as the continuous line pixel signal.
The result after correcting the slope is shown in thick lines. 
Note here that the signal of each pixel is not offseted
after the LTT correction 
(this is the flat-field correction detailed in Sect. \ref{FF}).
}
\end{figure}

Long Term Transients (LTT) still have an unknown origin. They are
observed on a characteristic time scale of about 60-80 minutes and are
the main sensitivity 
limitation of the ISOCAM diffuse emission analysis 
(Miville-Desch\^enes et al. 2000). LTT are observed in FIRBACK data 
(Fig. \ref{LTT_1}) but only for some rasters and pixels.
We first tried to correct the LTT using the method developped
by Miville-Desch\^ene et al. (2000) for ISOCAM;
however, this method fails since (1) ISOCAM and ISOPHOT detectors
are different and (2) the redundancy in our
observations is too low. The method developed for ISOCAM clearly detects 
the long term variation. However, for ISOPHOT, all detectors are not contaminated 
by the LTT (we see Fig. \ref{LTT_1} that the LTT is not seen 
by the four pixels), which is not the case
for ISOCAM, where all detectors have at a first order the same behaviour. 
Therefore, we correct the LTT by assuming that the
pixel, in each raster, that exhibits the smallest long term variation
(or smallest slope) is representative of the sky. Fig. \ref{LTT_1} shows the results
before and after the correction.

\subsection{\label{FF}Individual raster flat-fielding}
\begin{figure}
\epsfxsize=9.cm
\epsfbox{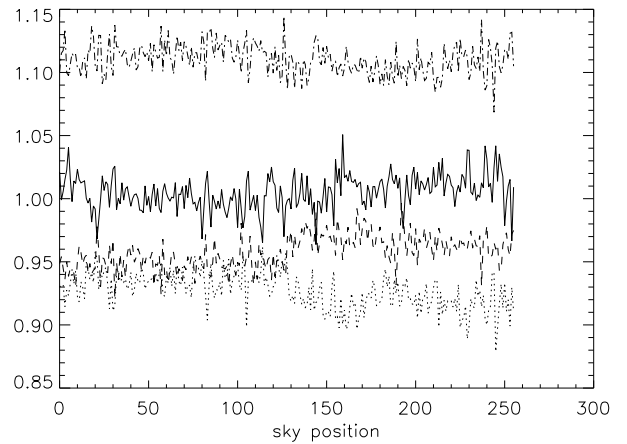}
\caption{\label{flat_1} Relative response of the 4 pixels as a function of
the sky position. The flat-field correction is obtained 
by computing a 30-points running median (and the mean correction is equal to 1).
}
\end{figure}

The flat-field correction, for each
individual raster, is directly estimated using
the redundancy of our observations and comparing
position by position the brightness of the 4 pixels.
First, for each raster, we compute the responsivity of the four pixels
with respect to the mean responsivity (Fig. \ref{flat_1}) for each
sky position. We see from Fig. \ref{flat_1} that each pixel has some high frequency 
instrumental noise. What we are interested in is the low-frequency difference
of responsivity between the four pixels. Therefore,
we derive the flat-field correction by computing a 30-points running median
(and the mean of the correction is equal to 1 per sky position).

The correction is rather constant for the whole 56 
rasters and equal on average to: $1.04 \pm 0.02$, $0.91 \pm 0.02$,
$1.09 \pm 0.02$ and $ 0.94 \pm 0.02$
for pixel 1, 2, 3 and 4 respectively. This highly reproductable
behaviour may be used to correct for the flat-fielding in any
other C-160 observation that has no redundancy.

\subsection{Transients induced by energetic cosmic rays}
The procedure described in Sect. 3.4 also reveals some long term transients
induced by energetic cosmic rays, as can been seen
in Fig. \ref{flat_1} for the dashed line signal
around the sky position number 130. 
These effects induced by energetic cosmic rays are
quite rare: we detect only 13 events for FIRBACK (to be compared to 
4 detectors $\times$ 56 rasters = 224 timelines).
After such a cosmic ray, the signal may
show an exponential-like behaviour (with a variable
time constant). We correct
for such events manually before computing the flat-field correction.

\subsection{\label{STT}Short term transients}
After a sudden increase of flux, most infrared detector 
signals shows an instantaneous jump followed by a slow rise
to the stabilisation level; this is the short term transient. 
In most observations, 
the stabilisation level is never reached, due to the limited
integration time per chopper plateau.\\

\begin{figure}
\epsfbox{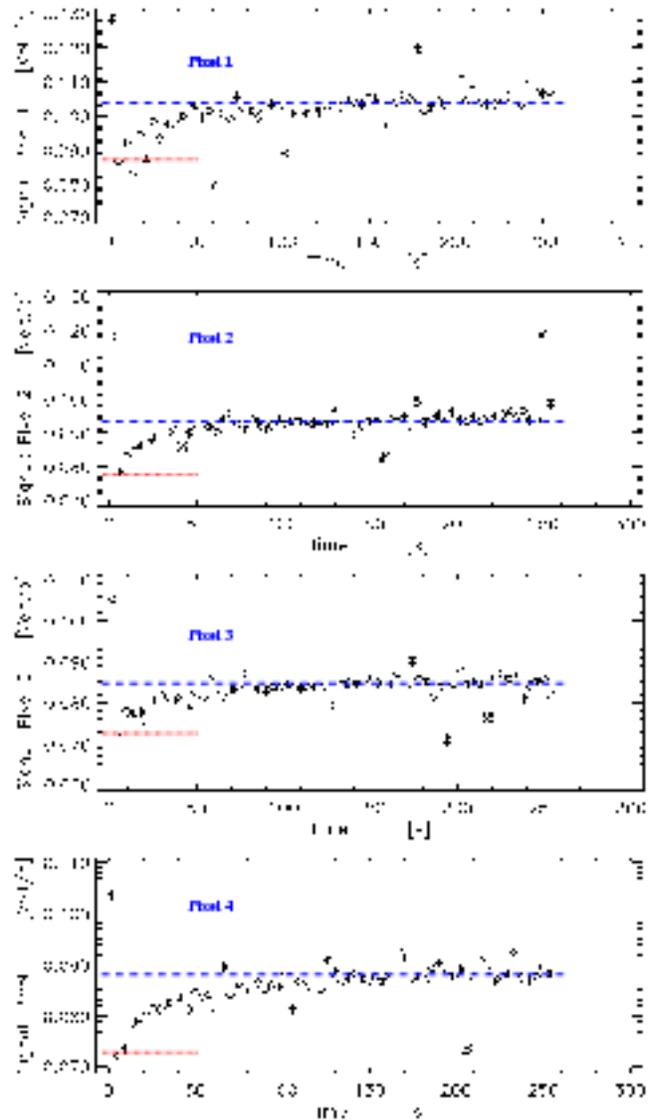}
\caption{\label{pht25_transient} Signal variation after a flux step (PHT25
sky measurements that follow the ``FCS dark'' measurements). 
For all pixels, the instantaneous jump is at the $\sim$85$\%$ level and the
stabilisation is reached in about 80s.}
\end{figure}

For the ISOPHOT C200 camera, we have information on the short 
term transients using our PHT25 absolute measurements. 
We see in Fig. \ref{pht25_transient} the signal behaviour of the PHT25
sky measurements for the 4 pixels. These measurements follow the ``FCS 
dark'' observation, whose signal is 10 times weaker (see Fig. \ref{pht25}). 
We clearly see in Fig. \ref{pht25_transient} a short term transient before 
the stabilisation. The instantaneous signal jump is of the order of 85$\%$; 
and the stabilisation is reached in several tens of seconds. 

%
\begin{figure*}[!ht]
	\begin{center}
	\epsfxsize=11cm
	\epsfbox{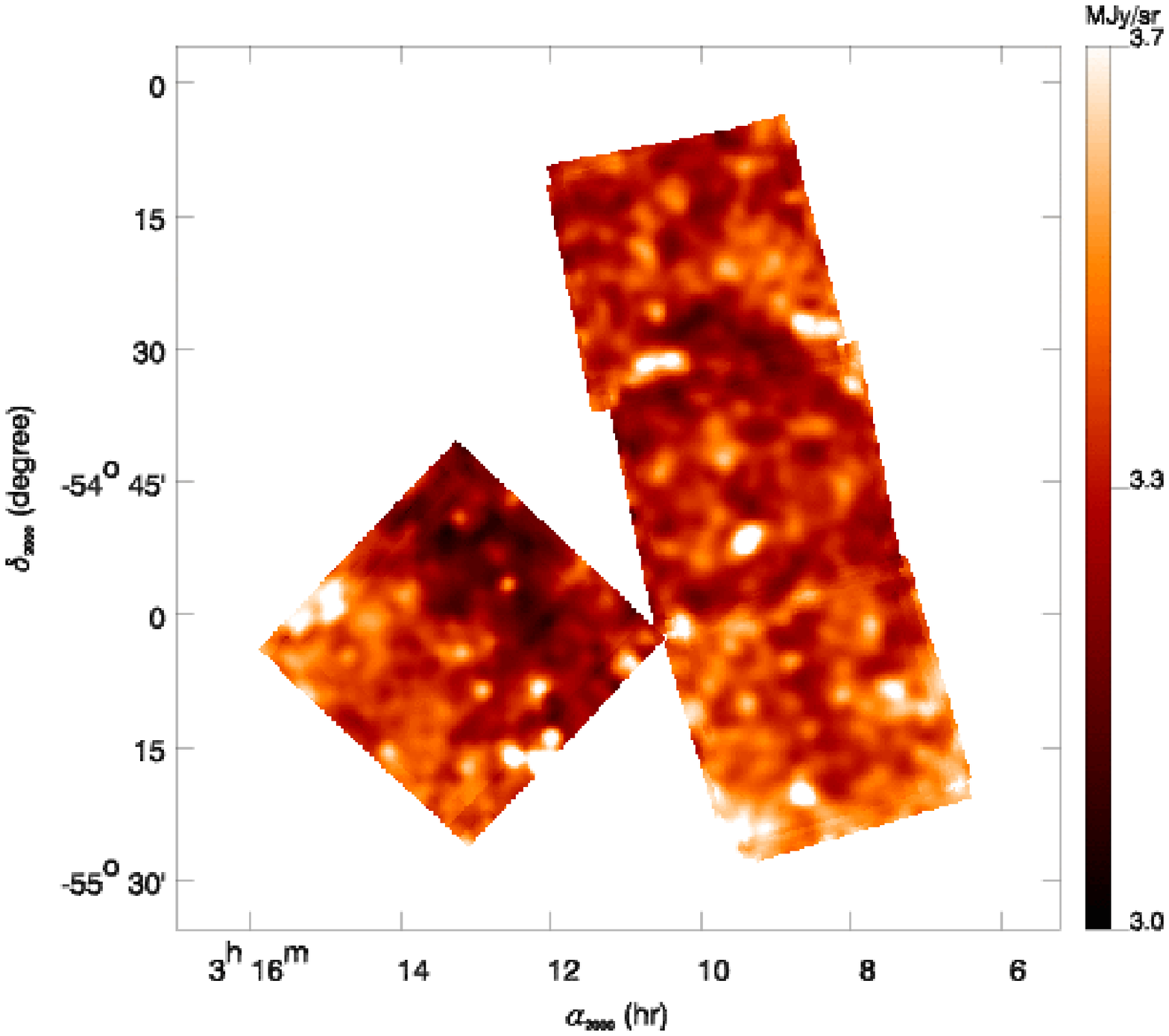}
	\caption[]{\label{Mar} Map of
	the FIRBACK South Marano (FSM) field.  FSM1 is the square on the left,
	and the rectangle is composed of  FSM2, 3 and 4 from top to bottom.}
	\end{center}
	\begin{center}
	\epsfxsize=15cm
	\epsfbox{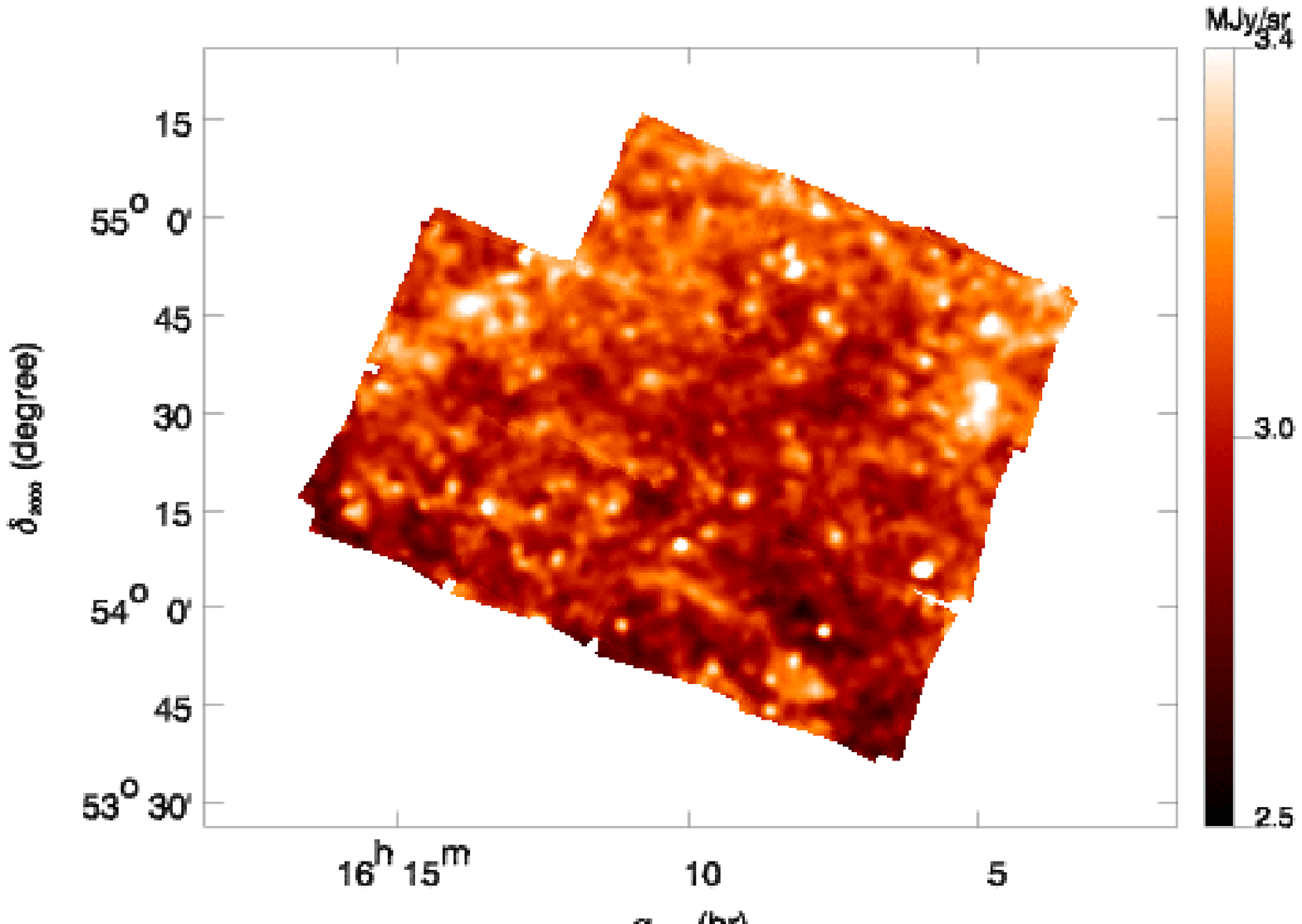}
	\caption[]{\label{N1} Map of the FIRBACK North 1 (FN1) field.}
	\end{center}
\end{figure*}
%
\begin{figure}[!ht]
	\begin{center}
	\epsfxsize=9cm
	\epsfbox{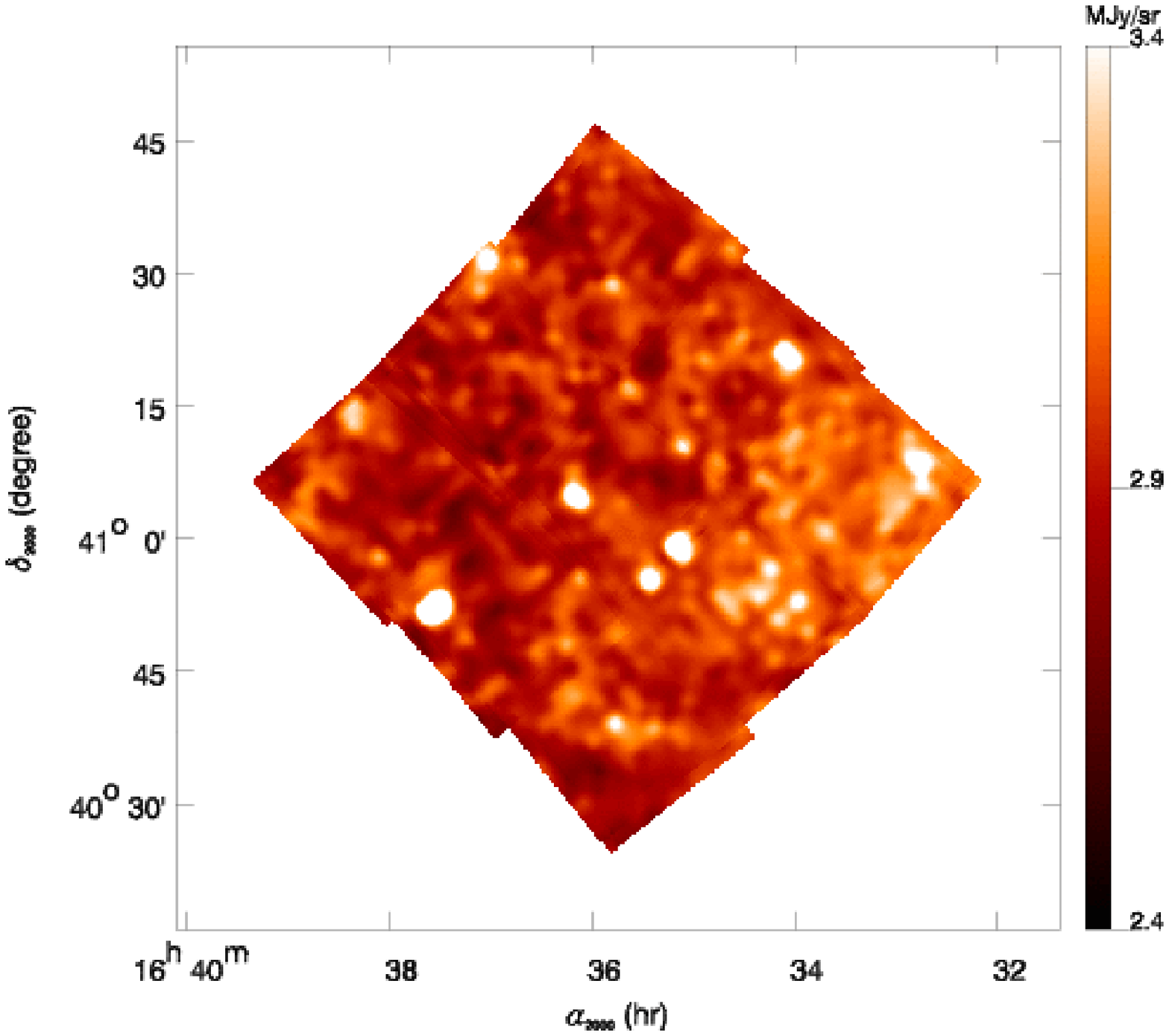}
	\caption[]{\label{N2} Map of the FIRBACK/ELAIS North 2 (FN2) field.}
	\end{center}
\end{figure}
%

For FIRBACK field observations, the short term transient effect is limited 
due to the flatness of the fields. Flux variations are 
very smooth and weak, except when point sources are observed. Therefore,
no correction is applied for the diffuse emission. In return, for point sources,
a correction of 10$\%$ is applied (this 10$\%$ is derived using the flux measured 
after 16s FIRBACK integration time compared to the stabilised flux).\\

For the FCS FIRBACK measurements, no correction is applied since (1) the difference
between the stabilised signal in the PHT25 FCS measurement and the one measured in 32s (which
is the integration time on the FCS) is less than 4 $\%$ and (2)
the second FCS observation is at the same level as the FIRBACK sky
observation (no flux step) and the two FCS measurements (before and after one observation)
differ by less than 3$\%$ for most of the rasters.

\subsection{\label{GFF}General ``flat-fielding''}
The different rasters in the same field have
been done at different time (there can be
several months between the rasters). Therefore,
the absolute calibration performed using
the FCS measurements can be slightly
different from one raster to another. This difference
is observed in the final FIRBACK maps as discontinuities
between individual rasters. \\

To correct for this small absolute calibration difference,
we apply a ``general flat-field'' correction in order to make
consistent the level of each raster with
its neighbourhood. This correction consists of
adding or subtracting offsets, the sum of the offsets
being 0. This effect is very small,
less than 5$\%$, except for one raster in FN1
which shows a discontinuity with its neighbourhood of 
about 11$\%$.

\subsection{\label{reproj}Reprojection}
The flux is finally projected on a 10"x10" coordinate 
grid using our own projection procedure. 
The procedure is the following:
\begin{itemize}
\item We compute each individual pixel coordinates.
\item We create individual maps for each pixel and raster
with 2D $\frac{sinx}{x}$ interpolation for the oversampling signal
and bilinear interpolation for the coordinates.
\item We compute for each field the final coordinate grid (10'' sampling)
using the individual raster coordinates 
\item At each final grid point, we add the observed signal and compute
the weight and variance.
\end{itemize}

We have checked the accuracy of our method using simulated sky maps,
containing both extended emission and point sources.
Our reprojection does not change the background brightness
and the sources flux (at the 1$\%$ level). One has, however, 
to note that this method is particularly well adapted to cosmological observations:
weak sources on a rather flat background (for fields with very bright
sources, this method may not preserve the point source photometry).\\

The final maps are shown in Fig. \ref{Mar}, \ref{N1} and \ref{N2}
for the three fields respectively.

\section{\label{cal} Calibration}

The extended emission calibration has changed 
by a factor of about 2 between PIA version 6 and 
PIA version 7. A large part of this factor comes 
from the different footprint solid angle values used 
in the different PIA versions. 

\subsection{\label{PSF} Footprint observations around Saturn}
\begin{figure}
\epsfxsize=9.cm
\epsfbox{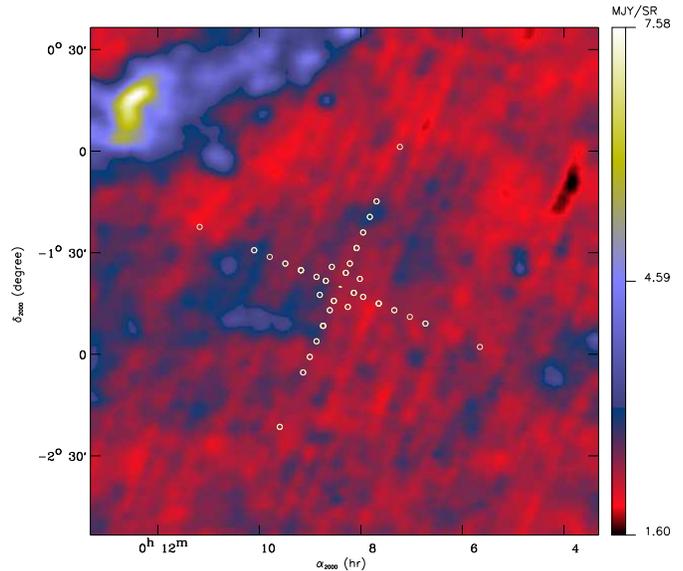}
\caption{\label{sat_1} ISOPHOT pointing positions (circle) around Saturn 
on the IRAS 100 $\mu$m map. Saturn positionnal changes during the observations 
have been computed using the ephemerids of the Bureau des Longitudes
(Berthier, 1998) 
and are represented by a small segment at the middle of the cross.}
\end{figure}

During the revolution 409, ISOPHOT made several tens of pointing
around Saturn at distances between 4.2 and 45.4 arcmin in
order to map the extended wings of the PHOT footprint at 170 $\mic$
during about 5 hours.
Observations were made in the Y and Z satellite axis directions.
Fig \ref{sat_1} shows the observed pattern on an IRAS
100 $\mu$m map. All directions were observed twice, back and forth.
Observations were performed with the following AOTs:
PHT 37-38-39 (sparse maps; PHT37: FCS; PHT38: sky; PHT39: FCS) 
for distances between 4.2 and 27.6 arcmin, and PHT25 (absolute photometry)
at the largest distance. Integration times are
around 150 sec for PHT38, 200 sec for PHT37-39 and
400 sec for PHT25. 

\begin{figure}
\begin{center}
\epsfxsize=9.cm
\vspace{2.5cm}
\hspace{1.cm}
\epsfbox{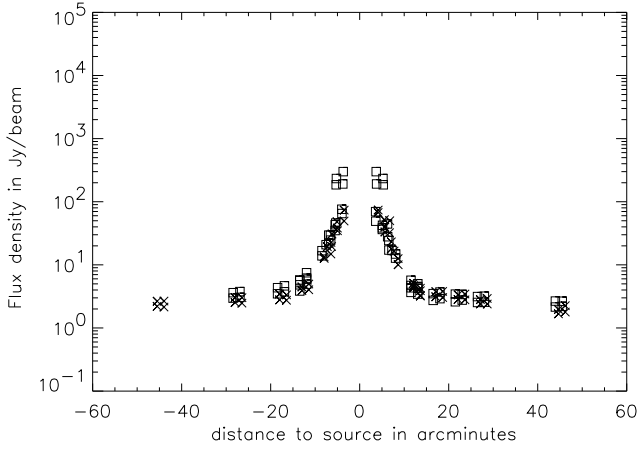}
\caption{\label{sat_2} Y direction measurements around Saturn. Squares are forward data, crosses, backward
data. Dispersion bewteen the 4 pixels at each position comes mainly
from the non correction of the flat-field.}
\end{center}
\end{figure}

\begin{figure}
\begin{center}
\epsfxsize=9.cm
\vspace{3.cm}
\hspace{1.cm}
\epsfbox{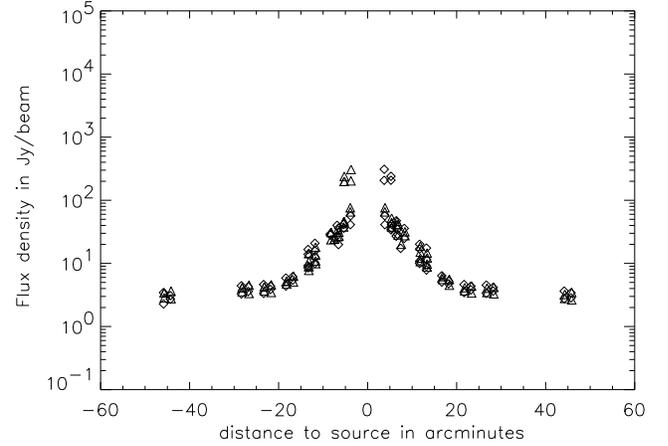}
\caption{\label{sat_3} Z direction measurements around Saturn. Triangles are forward data, diamonds, backward
data.}
\end{center}
\end{figure}

We use PIA V7.2.2 for the data reduction and calibration. For each detector and
each position, we keep only the second half of the data corresponding 
to the stabilised signal. These observations have no 
transients induced by cosmic rays and we do not apply
any flat-field correction. 
Sixty seven files were used, the others
being either unreadable (3 files) or saturated (3 files).
Results are presented in Fig. \ref{sat_2} and
\ref{sat_3} for all data in the Y and Z direction respectively. Each pixel
is plotted and is used as an independent measurement.

\subsection{Comparison of the footprint model with Saturn observations}
\begin{figure}
\epsfxsize=9.cm
\epsfbox{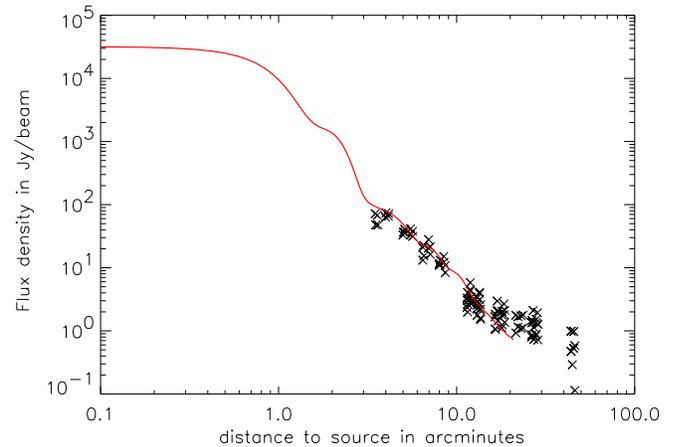}
\caption{\label{sat_4} ISOPHOT footprint model (continuous line) compared
to ISOPHOT measurements around Saturn for the Z axis.}
\end{figure}

A model for the ISOPHOT footprint has been developed at Heidelberg 
(Klaas et al., private communication) based on the ISOCAM footprint routines. This model includes
the optical characteristics of the telescope, the  primary and secondary mirrors,
and the filters and detectors. We have used this program to compute
the ISOPHOT footprint up to 20 arcmin with the 170~$\mu$m bandpass filter.
Results of the model are shown in Fig. \ref{sat_4} together with the
Saturn measurements for the Z axis.
To make this comparison, we have assumed
for Saturn a flux of 32000~Jy and removed the background
using the PHT25 measurement.

We see a very good agreement between the model and the measurements. 
However, around 4.5 arcmin from Saturn, some data, coming from one Y direction scan
(only one position), 
have a significantly higher flux than the model prediction. 
Data at similar distances from Saturn in other scans cannot be used due to 
saturation problems. Therefore, we interpret this discrepancy as due to 
detector saturation problems (one can note, however, that the contribution of these data points to the 
solid angle is lower than a few percent).

In conclusion, the ISOPHOT footprint measurements and model 
at 170 $\mu$m are in very good agreement. We therefore use, in the
following, the model (from Klaas et al., private communication) as the definitive footprint at 170 $\mic$.

\subsection{The extended brightness photometric correction factor}

In PIA V6.5 the solid angle used to convert the flux 
in brightness was that of the pixel (1.88 10$^{-7}$ sr);
in PIA V7.2.2, it is the footprint model's one
but truncated at 4.1 arcmin (2.73 10$^{-7}$ sr).
The full solid angle of the footprint is equal to 3.06 10$^{-7}$ sr.
Therefore, there is a photometric correction
factor to be applied to the extended emission:
$$B_{\nu} = B_{\nu}(piaV72) \times \frac{\Omega_{piaV72}}{\Omega_{footprint}}$$
$$B_{\nu} = B_{\nu}(piaV72) \times 0.89$$
After correction of the absolute calibration of the extended 
emission, point source fluxes (given in Dole et al. 2001)
are computed using the footprint model (Klaas et al., private communication).

\subsection{The rejection rate}
Unique measurements have been done by the ISOPHOT team during
the eclipse of the sun by the earth (Kranz et al. 1998; Klaas et al.
1998a; Lemke et al. 1998) by pointing at a sky region
at 60$^o$ from the sun before, during, and after the eclipse.
These measures reveal no signal variation, leading to an upper
limit of 10$^{-13}$ at 60$^o$ for the straylight rejection rate.
This exceptional measurement clearly shows that there is
no contribution to the flux from the far-side lobes.
This demonstrates that ISO is able to make absolute measurements 
of the extended emission and gives a high degree of confidence
to our absolute photometric calibration.

\subsection{Comparison with DIRBE extended brightness}
We can now compare the extended ISOPHOT FIRBACK brightness
with the predicted one using DIRBE and HI measurements.
Because of the size of the DIRBE beam,
this comparison is only feasible due to the flatness
of the FIRBACK fields; this flatness is observed with DIRBE
on several degrees around each FIRBACK field.\\

The sky measurement is the sum of the zodiacal light, Cosmic Infrared
Background (CIB) and dust interstellar emission.
For each field the zodical light, at the time of the observation,
has been computed using the Reach et al. (1995) DIRBE model.
The CIB is extrapolated at 170 $\mu$m using the 
DIRBE measurements of Lagache et al. (2000).
For the dust emission, we compute its contribution
using (1) the HI column density from the Leiden-Dwingeloo survey,
Hartmann \& Burton 1997 (we prefer to use the HI column density
rather than the DIRBE brightness since DIRBE data are very noisy in 
FIRBACK fields) and (2) the emissivity 
from Lagache et al. (2000). The final predicted emission
at 170 $\mu$m for the three fields is shown in
Table \ref{dirbe_compa}. It is in remarkable
agreement with the measured ISOPHOT brightness{\footnote{The difference between Puget et al. 
2000 (paper I) FSM1 brightness and the brightness obtained now is less than 6$\%$}}.\\

\begin{table}
\caption{Cosmic Infrared Background (CIB) from Lagache et al. (2000), zodiacal
(from Reach et al. 1995) and dust emission (from Lagache et al. 2000) at 170 $\mu$m 
for the three FIRBACK fields
(in MJy/sr). The total emission is the sum of the three contributions for each
field. There is a remarkable agreement between the predicted
brightness and the measured ISOPHOT one.}
\label{dirbe_compa}
\begin{tabular}{|l|c|c|c|} \hline 
 & FN1 & FN2 & FSM \\ \hline
CIB & 1.10$\pm$0.2 & 1.10$\pm$0.2 & 1.10$\pm$0.2 \\ \hline
Zodiacal & 0.71$\pm$0.1 & 0.80$\pm$0.1 & 0.75$\pm$0.1 \\ \hline
N$_{HI}$ (cm$^{-2}$) & 8.2 10$^{19}$ & 7.7 10$^{19}$ & 1.0 10$^{20}$ \\ \hline
Dust & 1.17$\pm$0.35 & 1.09$\pm$0.33 & 1.42$\pm$0.43 \\ \hline \hline
Total predicted & 2.98$\pm$0.41 & 2.99$\pm$0.40 & 3.27$\pm$0.48 \\ \hline \hline
PHOT measured & 3.01$\pm$0.14 & 2.97$\pm$0.17 & 3.39$\pm$0.12 \\ \hline
\end{tabular}\\
\end{table}

\begin{figure}
\epsfxsize=9.cm
\epsfysize=7.cm
\epsfbox{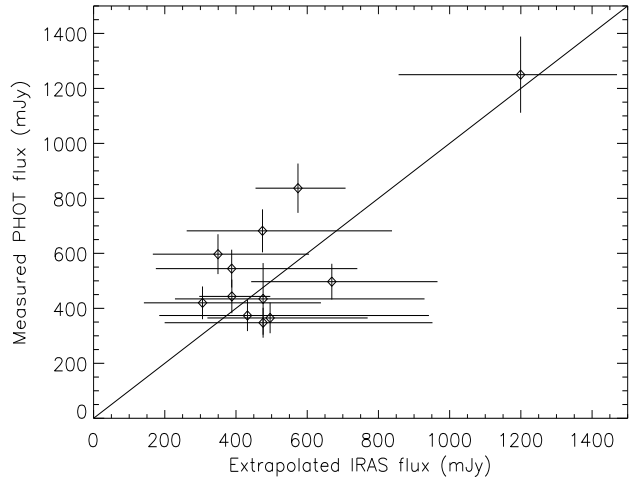}
\caption{\label{compa_IRAS} Comparison of the ISOPHOT 170 $\mu$m measured
flux with the IRAS extrapolated ones. Data are compatible with a slope
of unity.}
\end{figure}

\subsection{Comparison with IRAS point sources measurements}
The absolute flux calibration of ISOPHOT is derived using
calibration standards such as planets, asteroids or stars 
(Klaas et al. 1998b; Schulz et al. 1999; ISO Consortium 2000a, 2000b). 
One can however check, using IRAS detected sources in FIRBACK fields,
the consistency between IRAS 60 and 100~$\mu$m flux and the ISOPHOT 170 $\mu$m one.
However, we have to keep in mind that
it is very difficult to extrapolate the IRAS fluxes at 170 $\mu$m
(one can have a factor 2 to 3 in the extrapolation using different
models or black-body temperatures).\\

Twelve FIRBACK sources,
well identified as very nearby non interacting galaxies, have IRAS 60 and 100~$\mu$m
flux counterparts, measured with SCNAPI{\footnote{SCANPI is described on 
the WEB site $http://www.ipac.caltech.edu/ipac/iras/scanpi$\_$over.html$}}. 
We extrapolate the 60 and 100
$\mu$m IRAS flux using the template spectra of Dale et al. (2000). These spectra are
based on the IRAS 60/100 color ratio and are thus well adapted. The comparison
is shown on Fig. \ref{compa_IRAS}. Uncertainties on the extrapolation, 
that take into account only the errors on the 100 and 60 $\mu$m fluxes
(and not the model uncertainty), are very large. 
Data are compatible with a slope of unity; but
this comparison is only illustrative.

\section{Conlusions}
We have presented in this paper the final FIRBACK data 
reduction using: (1) The Phot Interactive Analysis version 7.2.2 and
(2) extra developments (corrections of the flat-field, 
long and short term transients, transients induced by cosmic rays and
adapted reprojection). Most of these extra developments
have been made possible by the perfect redundancy
inside each raster (one pixel overlap in both Y and Z direction).
We have then checked the absolute calibration
using PHT25 measurements and the footprint measured on Saturn
(compared to the footprint model). We have shown that the
ISOPHOT 170 $\mu$m extended emission calibration has to
be corrected by a factor 0.89, which comes from the difference
in solid angle between the PIA v7.2 and the modeled footprint.\\

Using this data reduction and calibration, we obtain an
absolute calibration which is in remarquable agreement
(better than 10$\%$) with brightness extrapolation that 
uses DIRBE data and HI column density measurements. 
We have in FIRBACK fields a very high signal to noise
ratio (greater than 50). The main limitation comes in fact from the
extragalactic source confusion itself (Dole et al. 2001).
\\

Acknowledgements: We would like to thank A. Abergel, C. Gabriel, U. Klaas, M-A.
Miville-Desch\^enes and the ISOPHOT team for many interactions
concerning the data reduction and the footprint analysis.
We thanks Danny Dale for his help in
using the template spectra for IRAS extrapolation. We thank J.L. Puget
for his helpful advice all along the data reduction process.

\end{document}